\documentclass[a4paper,nobibnotes,nofootinbib,preprintnumbers]{revtex4}
\usepackage{amssymb}
\usepackage{amsmath}
\usepackage{epsfig}
\newcommand{\be}{\begin{equation}}
\newcommand{\ee}{\end{equation}}
\newcommand{\bea}{\begin{eqnarray}}
\newcommand{\eea}{\end{eqnarray}}
\newcommand{\nn}{\nonumber}

\newcommand{\g}{\gamma}
\newcommand{\f}{\frac}

\newcommand{\bra}{\langle}
\newcommand{\ket}{\rangle}
\newcommand{\intc}[1]{{\int\frac{d#1}{2i\pi}}}
\newcommand\lr[1]{{\left({#1}\right)}}

\begin{document}
\title{Azimuthal decorrelation of Mueller-Navelet jets at the Tevatron and the LHC}
\author{C. Marquet}\email{marquet@quark.phy.bnl.gov}
\affiliation{RIKEN BNL Research Center, Brookhaven National Laboratory, Upton, NY 11973, USA}
\author{C. Royon}\email{royon@hep.saclay.cea.fr}
\affiliation{DAPNIA/Service de physique des particules, CEA/Saclay, 91191 
Gif-sur-Yvette cedex, France}
\preprint{RBRC-668}
\begin{abstract}

We study the production of Mueller-Navelet jets at hadron colliders in the Balitsky-Fadin-Kuraev-Lipatov (BFKL) framework. We show that a measurement of the relative azimuthal angle $\Delta\Phi$ between the jets can provide a good testing ground for corrections due to next-leading logarithms (NLL). Besides the well-known azimuthal decorrelation with increasing rapidity interval
$\Delta\eta$ between the jets, we propose to also measure this effect as a function of
$R=k_2/k_1,$ the ratio between the jets transverse momenta. Using renormalisation-group improved NLL kernel, we obtain predictions for $d\sigma/d\Delta\eta dR d\Delta\Phi.$ We analyse NLL-scheme and renormalisation-scale uncertainties, and energy-momentum conservation effects, in order to motivate a measurement at the Tevatron and the LHC.

\end{abstract}
\maketitle
\section{Introduction}

Mueller-Navelet jet production \cite{mnjets} in hadron-hadron scattering is a process in which a jet is detected in each of the forward directions with respect to the incident hadrons. This process is characterized by two hard scales: $k_1$ and $k_2,$ the transverse momenta of the forward jets. When the total energy of the collision 
$\sqrt{s}$ is sufficiently large, corresponding to a large rapidity interval between the jets $\Delta\eta\!\sim\!\ln(s/k_1k_2),$ Mueller-Navelet jet production is relevant for testing the Balitsky-Fadin-Kuraev-Lipatov (BFKL) approach \cite{bfkl}.

In fixed-order perturbative QCD calculations, the hard cross section is computed at fixed order with respect to $\alpha_s.$ The large logarithms coming from the strong ordering between the hadrons scale and the jets transverse momenta are resummed using the Dokshitzer-Gribov-Lipatov-Altarelli-Parisi (DGLAP) evolution equation
\cite{dglap} for the parton densities. However in the high-energy regime, other large logarithms arise in the hard cross section itself, due to the strong ordering between the energy $\sqrt{s}$ and the hard scales. These can be resummed using the BFKL equation, at leading (LL) and next-leading (NLL) logarithmic accuracy \cite{bfkl,nllbfkl}. 

On the phenomenological side, a first attempt to look for BFKL effects was performed
at the Tevatron (Run 1), using measurements of cross-section ratios (for same jet kinematics and two different center-of-mass energies squared $s$ and $\tilde{s}$) that are independent of the parton densities and allow to study more quantitatively the influence of the high-energy effects. The data \cite{mnjtev} overestimate the LL-BFKL prediction $(s/\tilde{s})^{4\bar\alpha\ln(2)},$ however it has been argued 
\cite{schmidt} that the measurement was biased by the use of upper $E_T-$cuts, the choice of equal lower $E_T-$cuts, and hadronization corrections. As a result, these tests on the relevance of the BFKL dynamics were not conclusive.

On the theoretical side, it was known that NLL corrections to the LL-BFKL predictions could be large due to the appearance of spurious singularities in contradiction with renormalization-group requirements. However it has been realised \cite{salam,CCS} that a renormalisation-group improved NLL regularisation can solve the singularity problem and lead to reasonable NLL-BFKL kernels (see also \cite{singnll} for different approaches). This motivates the present phenomenological study of NLL-BFKL effects in Mueller-Navelet jet production. Our analysis allows to study the NLL-BFKL framework, and the ambiguity corresponding to the dependence on the specific regularisation scheme. Our goal is to motivate further measurements at the Tevatron (Run 2) and at the LHC.

In Ref. \cite{nllf2} and \cite{nllfj}, such phenomenological investigations have been devoted to the proton structure function and forward-jet production in deep inelastic scattering. The NLL-BFKL effects were taken into account through an ``effective kernel'' (introduced in \cite{CCS}) using three different schemes (denoted S3 and S4 from \cite{salam} and CCS from \cite {CCS}). While for the structure function analysis the NLL corrections didn't really improve the BFKL description, it was definitively the case in the forward-jet analysis.

The present study is devoted to the $\Delta\Phi$ spectrum, where $\Delta\Phi$ is the relative azimuthal angle between the Mueller-Navelet jets. We implement the NLL-BFKL effects following \cite{nllf2} and \cite{nllfj}, using the S3 and S4 schemes. We study the magnitude of the NLL corrections with respect to the LL-BFKL results. We confirm the expectations \cite{kmmo} that those corrections slow down the azimuthal decorrelation with increasing $\Delta\eta.$ 

We propose to also investigate this effect as a function of $R=k_2/k_1,$ the ratio between the jets transverse momenta. This is inspired by the results of \cite{nllfj} which showed that NLL-BFKL corrections have more impact on the forward-jet cross-section when the measurement is sensitive to different values of (the forward-jet equivalent of) $R.$ We obtain predictions for
$d\sigma^{hh\!\to\!JXJ}/d\Delta\eta dR d\Delta\Phi$ and show that this would allow for a detailed study of the NLL-BFKL approach and the QCD dynamics of Mueller-Navelet jets.

The plan of the paper is the following. In section II, we present the phenomenological NLL-BFKL formulation of the Mueller-Navelet jet cross-section. In section III, we introduce the observable $d\sigma^{hh\!\to\!JXJ}/d\Delta\eta dR d\Delta\Phi$ relevant to study the $\Delta\Phi$ spectrum. In section IV, we present the predictions obtained using the S3 and S4 schemes and compare them with LL-BFKL predictions. We also discuss the dependence of our results with respect to the choice of the renormalization scale determining $\alpha_s,$ and we estimate the impact of energy-momentum conservation effects. Section V is devoted to conclusions and outlook.

\section{Mueller-Navelet jets in the NLL-BFKL framework}

Mueller-Navelet jet production in a hadron-hadron collision is represented in 
Fig.1 with the different kinematic variables. We denote $\sqrt{s}$ the total 
energy of the collision, $k_1$ and $k_2$ the transverse momenta of the two 
forward jets and $x_1$ and $x_2$ their longitudinal fraction of momentum with 
respect to the incident hadrons as indicated on the figure. 
$\Delta\Phi\!=\!\pi\!-\!\phi_1\!+\!\phi_2$ measures the relative azimuthal angle between the two jets, as $\phi_1$ and $\phi_2$ are the jets angles in the plane transerve to the collision axis. In the following, we 
consider the high-energy regime in which the rapidity interval between the 
two jets $\Delta\eta\!=\!\log(x_1x_2s/k_1k_2)$ is assumed to be very large. 
Following the phenomenological NLL-BFKL analysis of \cite{nllf2,nllfj}, one obtains 
the Mueller-Navelet jet cross section:
\be
\f{d\sigma^{hh\!\rightarrow\!JXJ}}{dx_1 dx_2 dk_1^2 dk_2^2 d\Delta\Phi}=
\f{\alpha_s(k_1^2)\alpha_s(k_2^2)}{4k_1^4k_2^2}f_{eff}(x_1,k_1^2)f_{eff}(x_2,k_2^2)
\sum_{p=-\infty}^\infty\intc{\g}\lr{\f{k_1^2}{k_2^2}}^\g
\ e^{\bar\alpha(k_1k_2)\chi_{eff}[p,\g,\bar\alpha(k_1k_2)]
\Delta\eta+ip\Delta\Phi}\label{mnjnll}
\ee
with the complex integral running along the imaginary axis from $1/2\!-\!i\infty$ 
to $1/2\!+\!i\infty.$ The running coupling is
\be
\bar\alpha(k^2)=\alpha_s(k^2)N_c/\pi=
\left[b\log\lr{k^2/\Lambda_{QCD}^2}\right]^{-1}\ ,
\hspace{1cm}b=\f{11N_c-2N_f}{12N_c}\ .\label{runc}\ee

Let us give some more details on formula \eqref{mnjnll}.

\begin{itemize}

\item The NLL-BFKL effects are phenomenologically taken into account by the 
effective kernels $\chi_{eff}(p,\g,\bar\alpha).$ For $p=0,$ the scheme-dependent 
NLL-BFKL kernels provided by the regularisation procedure $\chi_{NLL}\lr{\g,\omega}$ 
depend on $\g,$ the Mellin variable conjugate to $k^2_1/k^2_2$ and $\omega,$ the 
Mellin variable conjugate to $s/s_0$ where $s_0=k_1k_2$ is the energy scale. In each case, the NLL kernels obey a {\it consistency condition} \cite{salam} which allows to 
reformulate the problem in terms of $\chi_{eff}(\g,\bar\alpha).$ The effective 
kernel $\chi_{eff}(\g,\bar\alpha)$ is obtained from the NLL kernel 
$\chi_{NLL}\lr{\g,\omega}$ by solving the implicit equation
$\chi_{eff}=\chi_{NLL}\lr{\g,\bar\alpha\ \chi_{eff}}$ as a solution of the
consistency condition. 

In the case of the S3 and S4 schemes \cite{salam} (in which $\chi_{NLL}$ is 
supplemented by an explicit $\bar\alpha$ dependence), we will extend the
regularisation procedure to non zero conformal spins and obtain 
$\chi_{NLL}\lr{p,\g,\omega};$ this is done in the Appendix. Then the effective kernels 
$\chi_{eff}(p,\g,\bar\alpha)$ are obtained from the NLL kernel by solving the implicit equation:
\be
\chi_{eff}=\chi_{NLL}\lr{p,\g,\bar\alpha\ 
\chi_{eff}}\ .
\label{eff}
\ee

\item In formula \eqref{mnjnll}, the renormalisation scale determinig $\bar\alpha$ is $k^2\!=\!k_1k_2,$ in agreement with the energy scale $s_0$ \cite{renscal,modrs}. In Section IV, we shall test the sensitivity of our results when using 
$k^2\!=\!\lambda\ k_1k_2$ and varying $\lambda.$ This is done using formula \eqref{mnjnll} with the appropriate substitution \cite{nllfj}
\be
\bar\alpha(k_1k_2)\!\rightarrow\!
\bar\alpha(\lambda k_1k_2)\!+\!b\ \bar\alpha^2(k_1k_2)\log(\lambda)\ ,
\ee
and with the effective kernel modified accordingly following formula 
\eqref{eff}. We also modify the energy scale into $s_0\!=\!\lambda\ k_1k_2.$

\item It is important to note that in formula \eqref{mnjnll}, we used the 
leading-order (Mellin-transformed) impact factors. We point out that 
the next-leading impact factors are known \cite{ifnlo}, and that in principle,
a full NLL analysis of Mueller-Navelet jets is feasible, but this goes
beyond the scope of our study. Also, our formula is different from the one 
proposed in \cite{sabsch}, because the authors considered the cross-section integrated with respect to the jets transverse momenta. This leads to a modification
of the jet impact factors which results in an extra factor $\g^{-1}(1-\g)^{-1}$
in the integrand of \eqref{mnjnll}. Also it modifies the effective kernel (see
\cite{sabsch} where the S3 scheme was considered).

\item In formula \eqref{mnjnll}, $f_{eff}(x,k^2)$ is the effective parton 
distribution function and resums the leading logarithms 
$\log(k^2/\Lambda_{QCD}^2).$ It has the following expression
\be
f_{eff}(x,k^2)=g(x,k^2)+\f{C_F}{N_c}\lr{q(x,k^2)+\bar{q}(x,k^2)}\ ,
\label{sf}\ee
where $g$ (resp. $q$, $\bar{q}$) is the gluon (resp. quark, antiquark) 
distribution function in the incident proton. Since the Mueller-Navelet jet measurement 
involves perturbative values of $k_1$ and $k_2$ and moderate values of $x_1$ and
$x_2,$ formula \eqref{mnjnll} features the collinear factorization of $f_{eff},$ 
with $k_1^2$ and $k_2^2$ chosen as factorization scales.

\end{itemize}

\begin{figure}[t]
\begin{center}
\epsfig{file=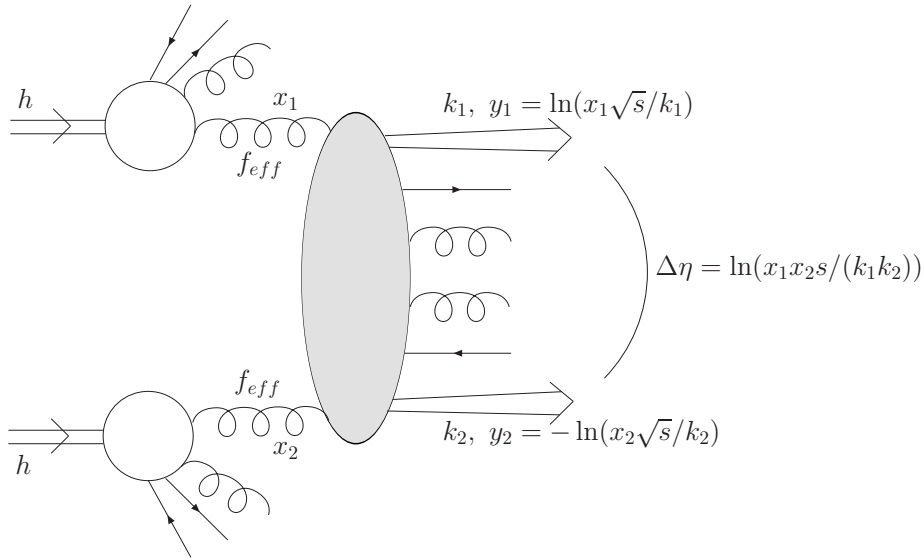,width=12cm}
\caption{Mueller-Navelet jet production in a hadron-hadron collision. The 
kinematic variables of the problem are displayed. $s$ is the total energy 
squared, $k_1$ ($y_1$) and $k_2$ ($y_2$) are the transverse momenta (rapidities) of the jets and $x_1$ and $x_2$ are their longitudinal momentum fraction with respect to the incident hadrons. $\Delta\eta$ is the rapidity interval between the hard probes.}
\end{center}
\end{figure}

By comparison, the LL-BFKL formula is formally the same as \eqref{mnjnll}, 
with the substitutions
\be
\chi_{eff}(p,\g,\bar\alpha)\rightarrow\chi_{LL}(p,\g)
=2\psi(1)-\psi\lr{1-\g+\f{|p|}2}-\psi\lr{\g+\f{|p|}2}\ ,
\hspace{1cm}\bar\alpha(k^2)\rightarrow\bar\alpha=\mbox{const. parameter}\ ,
\label{chill}\ee
where $\psi(\g)\!=\!d\log\Gamma(\g)/d\g$ is the logarithmic derivative of the 
Gamma function.

\section{The $\Delta\Phi$ spectrum}

\begin{figure}[t]
\begin{center}
\epsfig{file=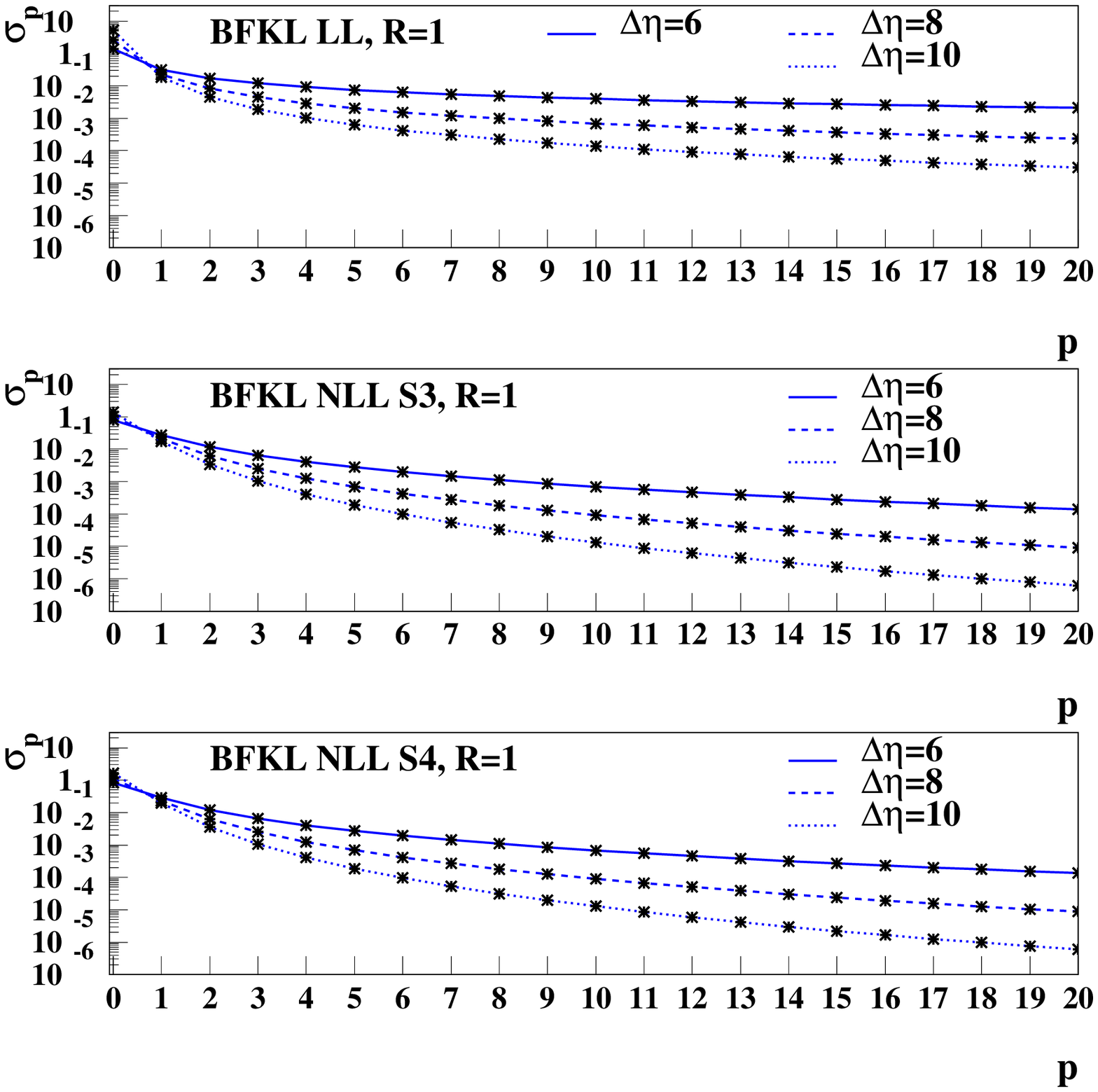,width=8cm}
\hfill
\epsfig{file=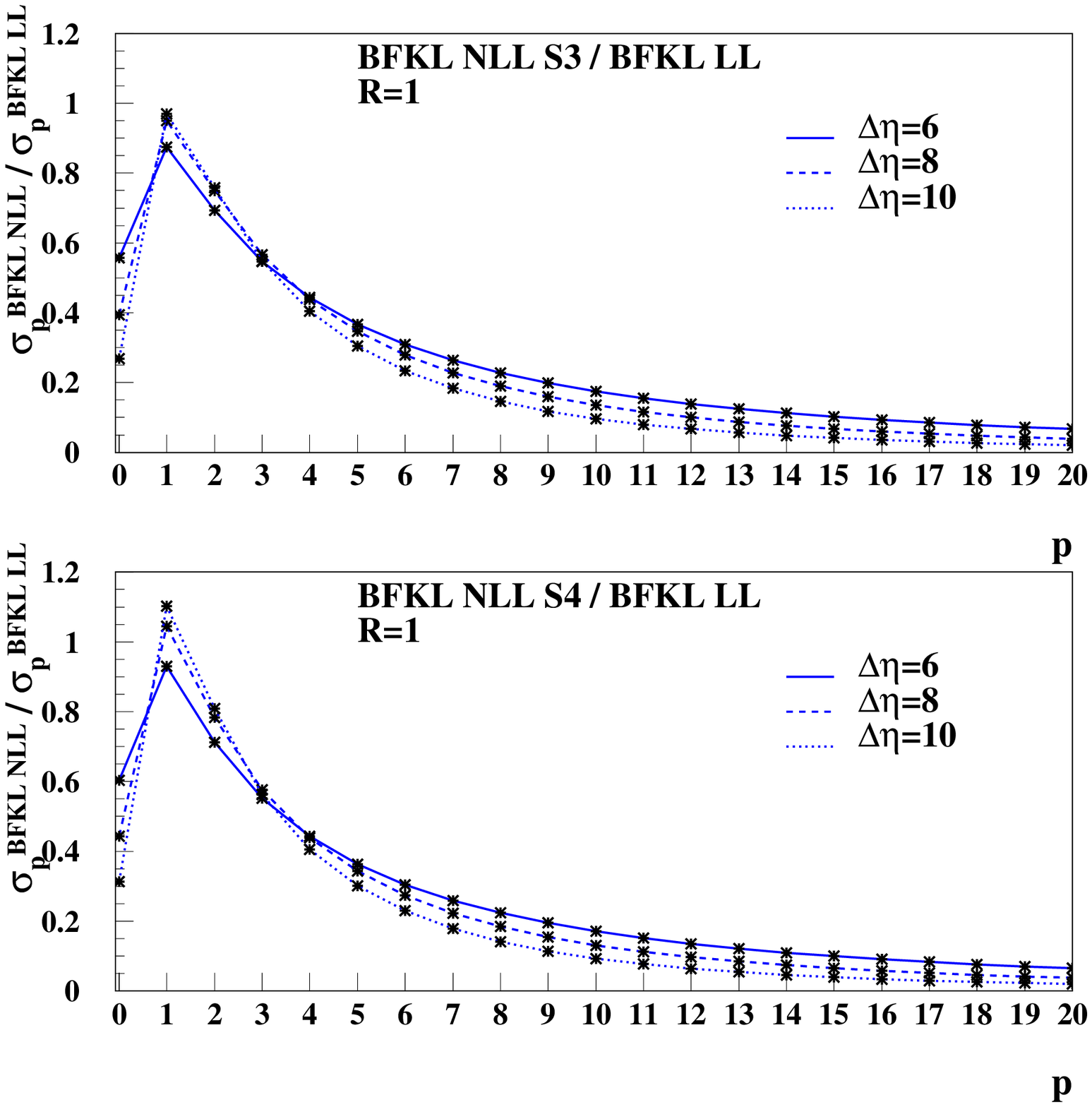,width=8cm}
\caption{Left plots: values of $\tilde\sigma_p(\Delta\eta,R\!=\!1)$ (see formula 
\eqref{tilsig}) entering into the $\Delta\Phi$ spectrum for the rapidity intervals 
$\Delta\eta=6,\ 8,\ 10;$ upper plot: LL-BFKL, middle plot: S3 scheme, lower plot:
S4 scheme. Right plots: ratios $\tilde\sigma_p^{NLL}/\tilde\sigma_p^{LL};$ upper plot: S3 scheme/LL-BFKL, lower plot: S4 scheme/LL-BFKL.}
\end{center}
\end{figure}

We would like to study the azimuthal decorrelation of the Mueller-Navelet jets as a function of their transverse momenta $k_1$ and $k_2$ and rapidities $y_1$ and $y_2:$
\be
y_1=\log\lr{\f{x_1\sqrt{s}}{k_1}}\ ,\hspace{1cm}
y_2=-\log\lr{\f{x_2\sqrt{s}}{k_2}}\ .
\label{raps}
\ee
Let us first introduce kinematic variables suitable for our problem: we change the variables in \eqref{mnjnll} to the variables
\be
\Delta\eta=y_1-y_2\ ,\hspace{1cm}y=\f{y_1+y_2}2\ ,\hspace{1cm}
Q=\sqrt{k_1k_2}\ ,\hspace{1cm}\mbox{and }R=\f{k_2}{k_1}\ .
\ee
One obtains
\bea
\f{d\sigma^{hh\!\rightarrow\!JXJ}}{d\Delta\eta dy dQ dR d\Delta\Phi}=
\f{\alpha_s(Q^2/R)\alpha_s(Q^2R)}{Q^3} x_1f_{eff}(x_1,Q^2/R)x_2f_{eff}(x_2,Q^2R)
\nn\\\sum_{p=-\infty}^\infty\int_{1/2-\infty}^{1/2+\infty}\f{d\g}{2i\pi}
R^{-2\g}\ e^{\bar\alpha(Q^2)\chi_{eff}[p,\g,\bar\alpha(Q^2)]\Delta\eta+ip\Delta\Phi}\ .
\label{newvar}\eea

We are interested in the following observable, suitable to study the azimuthal decorrelation of the jets as a function of their rapidity separation $\Delta\eta$ and of the ratio of their transverse momenta $R:$
\be
2\pi\left.\f{d\sigma}{d\Delta\eta dR d\Delta\Phi}\right/\f{d\sigma}{d\Delta\eta dR}=
1+\f{2}{\sigma_0(\Delta\eta,R)}\sum_{p=1}^\infty \sigma_p(\Delta\eta,R) 
\cos(p\Delta\Phi)\ .
\label{obs}\ee
We have expressed the normalized cross-section \eqref{obs} in terms of the Fourier coefficients
\be
\bra\cos(p\Delta\Phi)\ket=
\lr{\f{d\sigma}{d\Delta\eta dR}}^{-1}
\int d\Delta\Phi \cos(p\Delta\Phi) \f{d\sigma}{d\Delta\eta dR d\Delta\Phi}=
\f{\sigma_p(\Delta\eta,R)}{\sigma_0(\Delta\eta,R)}\ee
with the cross-sections $\sigma_p(\Delta\eta,R)$ obtained from \eqref{newvar} and given by
\bea
\sigma_p(\Delta\eta,R)= \int_{E_T}^\infty \f{dQ}{Q^3}\alpha_s(Q^2/R)\alpha_s(Q^2R)
\lr{\int_{y_<}^{y_>} dy\ x_1f_{eff}(x_1,Q^2/R)x_2f_{eff}(x_2,Q^2R)}
\nn\\\times\int_{1/2-\infty}^{1/2+\infty}\f{d\g}{2i\pi}R^{-2\g}
\ e^{\bar\alpha(Q^2)\chi_{eff}[p,\g,\bar\alpha(Q^2)]\Delta\eta}\ .
\label{coeff}\eea
The kinematical cuts $Q>E_T$ and $y_< < y < y_>$ for the $Q$ and $y$ integrations in \eqref{coeff} will be specified later, when we discuss the Tevatron and LHC kinematical ranges.

For the sake of comparison between BFKL LL and NLL effects, we define the following quantities, free of parton distribution functions:
\be
\tilde\sigma_p(\Delta\eta,R,\bar\alpha)=
\int_{1/2-\infty}^{1/2+\infty}\f{d\g}{2i\pi}R^{-2\g}
\ e^{\bar\alpha\chi_{eff}[p,\g,\bar\alpha]\Delta\eta}\ .
\label{tilsig}\ee
Note that in the LL-BFKL case in which $\bar\alpha$ does not depend on $Q^2,$ one has $\tilde\sigma_p/\tilde\sigma_0=\sigma_p/\sigma_0.$ We shall compare the LL and NLL values of $\tilde\sigma_p(\Delta\eta,R,0.16)$ for $R=1$ and $\Delta\eta=6,\ 8,\ 10.$ 
The comparison is shown on Fig.2 where we consider both the S3 and S4 NLL schemes. 

The cross sections $\tilde\sigma_p$ are displayed as a function of $p$ and, as expected for the rather large values of $\Delta\eta$ considered, we see that $\tilde\sigma_0$ is the largest cross section, and its increase with rapidity is stronger at LL compared to NLL. For $p\!\neq\!0,$ $\tilde\sigma_p$ decreases as a function of $\Delta\eta,$ and the ratios $\tilde\sigma_p^{NLL}/\tilde\sigma_p^{LL}$ between the NLL and LL contributions show that the decrease is faster at NLL except for $p=1$ and $p=2$ (and for $p=3$ the rapidity dependences at LL and NLL are comparable).

\section{Results for Mueller-Navelet jet $\Delta\Phi$ distributions}

\begin{figure}[t]
\begin{center}
\epsfig{file=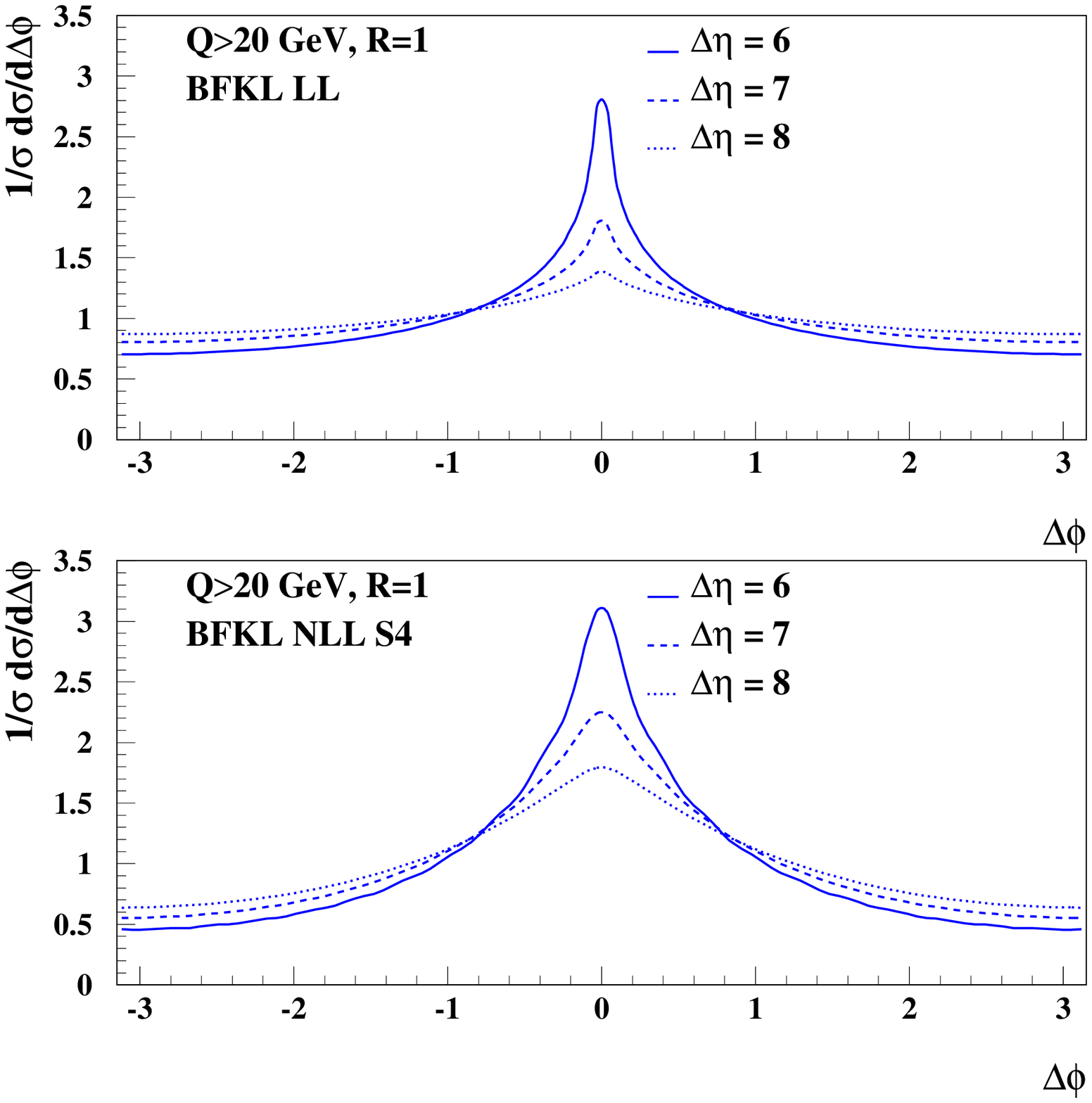,width=8.3cm}
\hfill
\epsfig{file=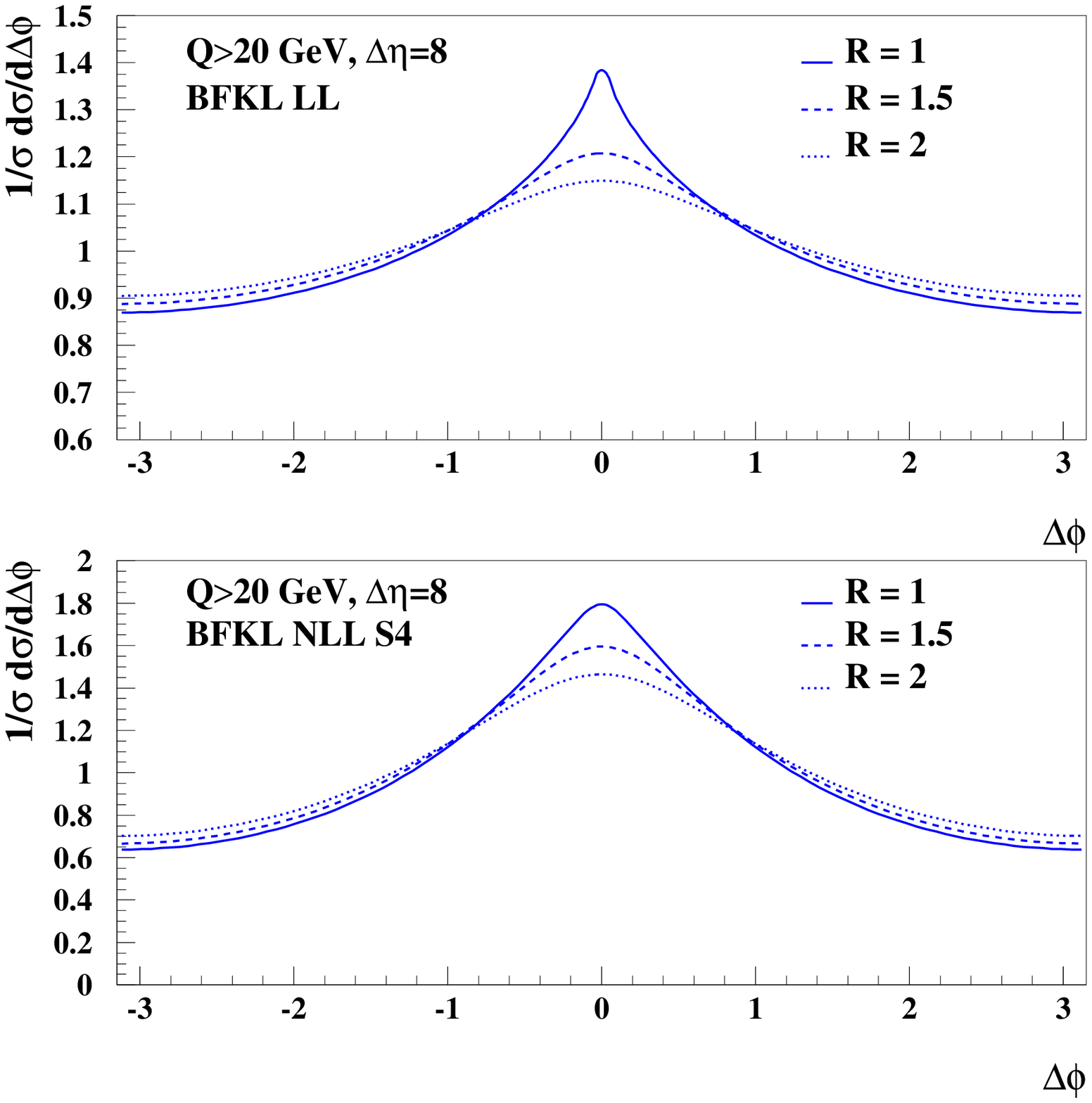,width=8.3cm}
\caption{The Mueller-Navelet jet $\Delta\Phi$ distribution \eqref{obs} for Tevatron (run 2) kinematics in the BFKL framework at LL (upper plots) and NLL-S4 (lower plots) accuracy. Left plots: $R=1$ and $\Delta\eta=6,\ 7,\ 8.$ Right plots: $\Delta\eta=8$ and $R=1,\ 1.5,\ 2.$}
\end{center}
\end{figure}

In this section, we show the results for the $\Delta\Phi$ distribution obtained with formulae \eqref{obs} and \eqref{coeff}. As shown in Fig.2, $\tilde\sigma_p$ decreases as a function of $p,$ and the decrease is faster at NLL compared to LL (and is similar for both schemes S3 and S4). As a result, including 20 terms in the sum over $p$ in \eqref{obs} is enough in the S3 and S4 cases. However at LL, one has to include more terms depending on the value of $\Delta\eta$ and $R.$

We choose to apply the rapidity cut $|y|<0.5$ which enforces a symmetric situation 
$y_2\!\sim\!-y_1.$ For the transverse momentum cut $E_T,$ we will consider two options corresponding to the Tevatron and the LHC possibilities in terms of kinematical reach:
$E_T\!=\!20\ \mbox{GeV}$ for the Tevatron (Run 2) and $E_T\!=\!50\ \mbox{GeV}$ for the LHC. We recall that the respective center-of-mass energies are 
$\sqrt{s}\!=\!1960\ \mbox{GeV}$ and $\sqrt{s}\!=\!14\ \mbox{TeV}.$

We point out that our NLL-BFKL predictions for the observable \eqref{obs} are parameter free. In the LL-BFKL case that we consider for comparisons, the only parameter $\bar\alpha$ is fixed to the value 0.16 obtained in \cite{llfj} by fitting on forward-jet data from HERA. By contrast, in the NLL-BFKL case, the value of 
$\bar\alpha$ is imposed by the renormalisation group equations.

\subsection{Comparison between LL and NLL BFKL predictions at the Tevatron and the LHC}

In Fig.3 and Fig.4, we display the observable \eqref{obs} as a function
of $\Delta\Phi$, for Tevatron and LHC kinematics respectively. The results are displayed for different values of $\Delta\eta$ and $R$ and at both LL and NLL accuracy (in this case, the S4 scheme is used). In general, the $\Delta\Phi$ spectra are peaked around $\Delta\Phi\!=\!0,$ which is indicative of jet emissions occuring back-to-back. 
In addition the $\Delta\Phi$ distribution flattens with increasing 
$\Delta\eta\!=\!y_1\!-\!y_2$ or with $R\!=\!k_2/k_1$ deviating from 1. Note the change of scale on the vertical axis which indicates the magnitude of the NLL corrections with respect to the LL-BFKL results. The NLL corrections slow down the azimuthal angle decorrelations for both increasing $\Delta\eta$ and $R$ deviating from $1.$

In the BFKL framework, the $\Delta\Phi$ dependence of the spectrum \eqref{obs} is larger at NLL than at LL. However, this $\Delta\Phi$ dependence is still smaller than in the fixed-order pQCD approach, in which the back-to-back peak is quite pronounced. Therefore a measurement of the cross-section $d\sigma^{hh\!\to\!JXJ}/d\Delta\eta dR d\Delta\Phi$ at the Tevatron (Run 2) or the LHC would allow for a detailed study of the QCD dynamics of Mueller-Navelet jets. In particular, measurements with values of $\Delta\eta$ reaching 8 or 10 will be of great interest, as these could allow to distinguish between BFKL and DGLAP resummation effects and would provide important tests for the relevance of the BFKL formalism. In addition, measuring the normalized cross-section \eqref{obs} could help reducing the biases which altered previous measurements 
\cite{mnjtev,schmidt}.

The D0 collaboration at the Tevatron (Run 1) did measure the azimuthal angle distribution between two jets \cite{azitev}, but they were not separated in rapidity by more than 5 units, in which case we do not expect the BFKL predictions to be relevant. Nevertheless, fixed order QCD predictions at next-to-leading order failed to describe the data, underestimating the decorrelation. In contrast, NLL-BFKL calculations overestimate the decorrelation \cite{sabsch}. Solving this puzzle likely requires to measure Mueller-Navelet jets with higher values of $\Delta\eta.$

\begin{figure}[t]
\begin{center}
\epsfig{file=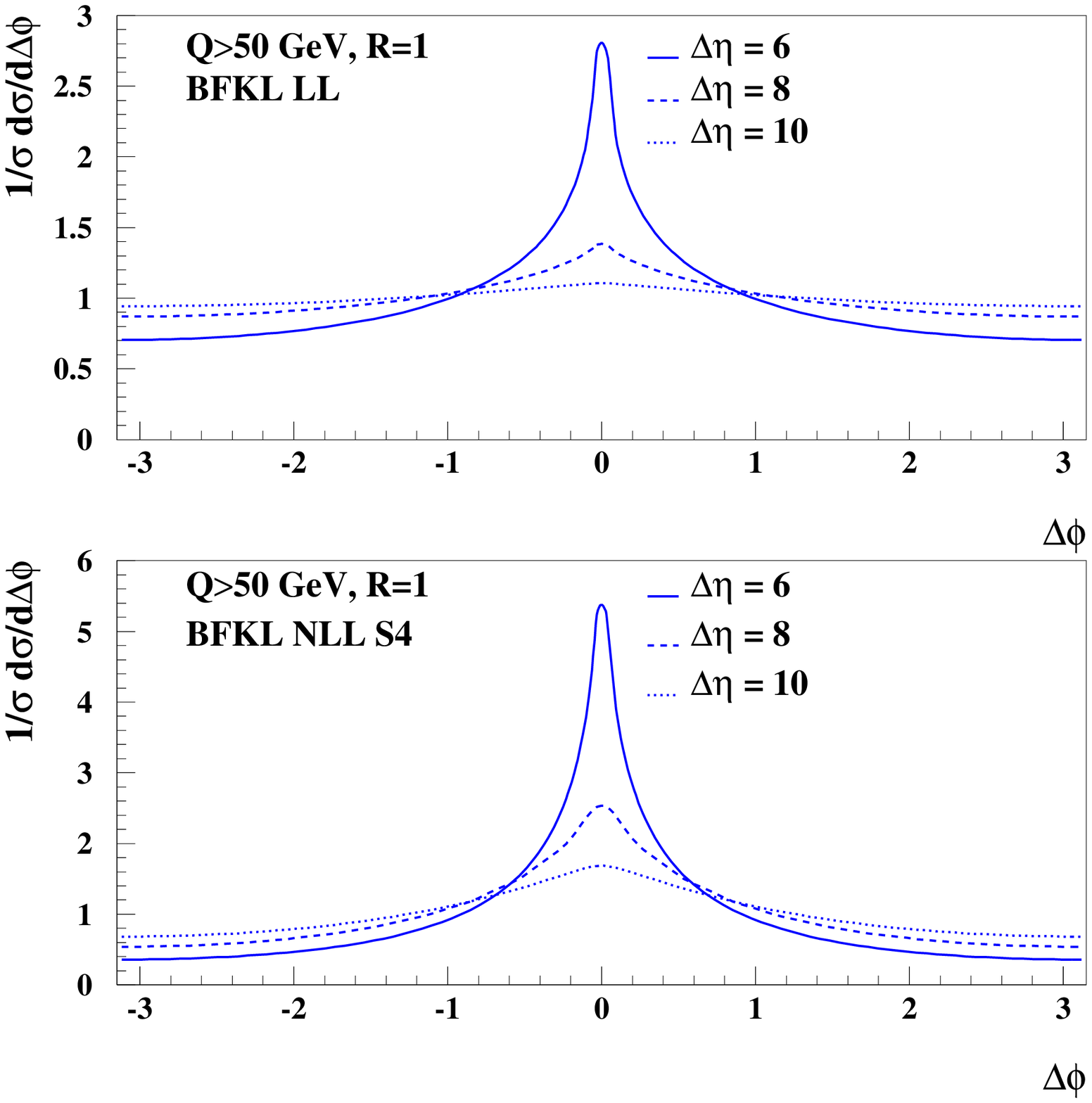,width=8.3cm}
\hfill
\epsfig{file=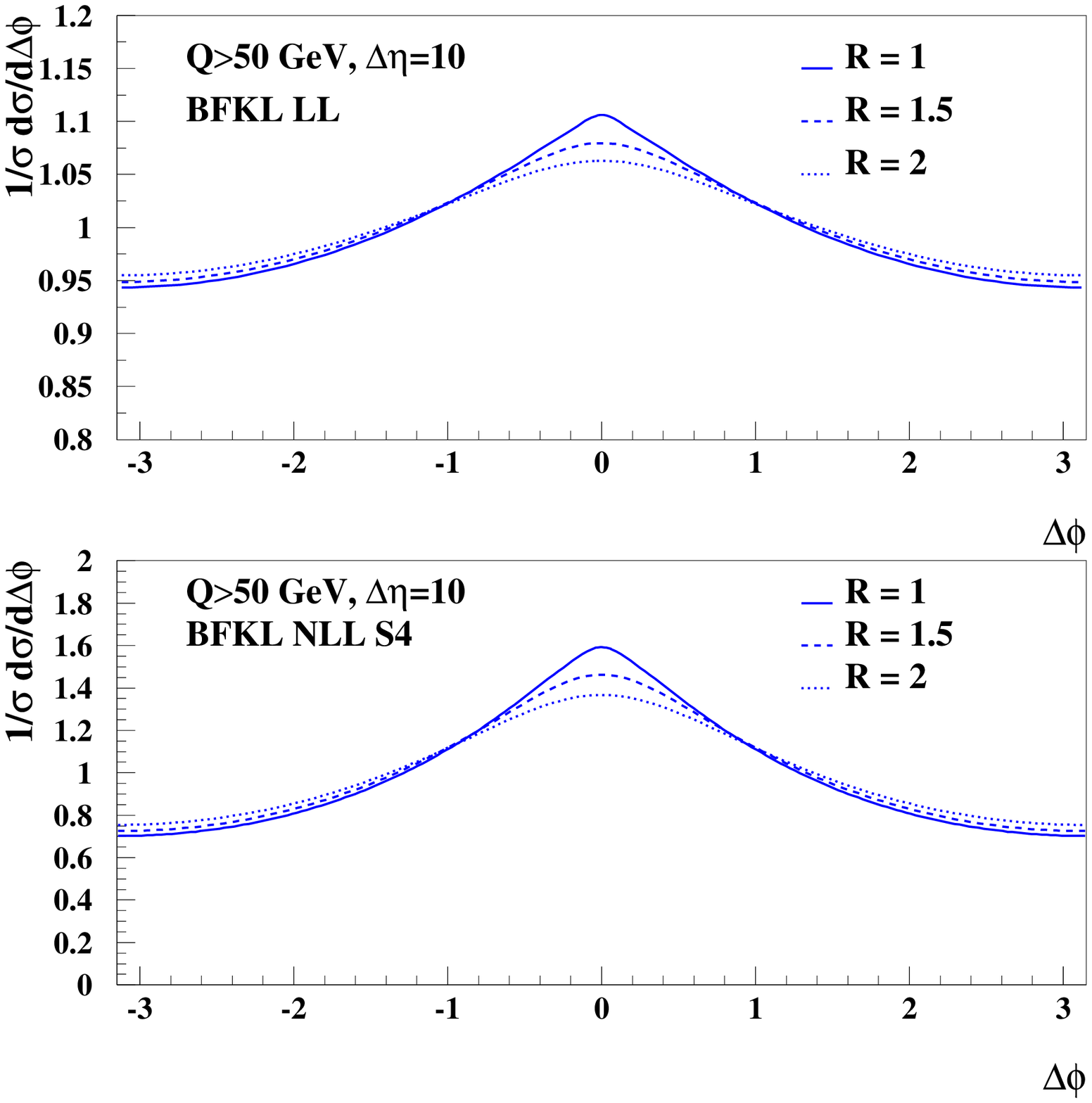,width=8.3cm}
\caption{The Mueller-Navelet jet $\Delta\Phi$ distribution \eqref{obs} for LHC kinematics in the BFKL framework at LL (upper plots) and NLL-S4 (lower plots) accuracy. Left plots: $R=1$ and $\Delta\eta=6,\ 8,\ 10.$ Right plots: $\Delta\eta=10$ and $R=1,\ 1.5,\ 2.$}
\end{center}
\end{figure}

\subsection{Scheme and scale dependence}

Our previous results in the NLL-BFKL case were obtained with the S4 scheme. As shown in Fig.2, the S3 scheme leads to similar results for the quantities  
$\tilde\sigma_p(\Delta\eta,R)$ and this is also true for the cross-sections 
$\sigma_p(\Delta\eta,R)$ (formula \eqref{coeff}) that actually enter in the formulation of the observable \eqref{obs}. There are some differences between the S3 and S4 scheme, but they tend to cancel when computing the ratios $\sigma_p/\sigma_0$ to obtain the $\Delta\Phi$ spectrum. Therefore the results obtained with both schemes are almost not distinguishible, as displayed on the left plots of Fig.5. Let us also point out that the pdf uncertainties cancel in the same way, and that the effects (not implemented here) due to the next-to-leading order jet impact factors would be suppressed too.

Let us now study the renormalization scale dependence of the NLL-BFKL description of Mueller-Navelet jets. Previously, the choice was $k_1k_2\!=\!Q^2$ and we now 
test the sensitivity of our results when using $Q^2/2$, and $2Q^2.$ We use formula \eqref{mnjnll} with the appropriate substitution $\bar\alpha(Q^2)\!\rightarrow\!
\bar\alpha(\lambda Q^2)\!+\!b\ \bar\alpha^2(Q^2)\log(\lambda)$ and with the effective kernel modified accordingly following formula \eqref{eff}. We also modify the energy scale $Q^2\!\rightarrow\!\lambda\ Q^2.$ The results are shown on the right plots of Fig.5, and the dependence on the choice of scale turns out to be quite small, of about 5 percent, except for $\Delta\Phi$ close to 0, in which case the uncertainty reaches 20 percent.

\subsection{Energy-momentum conservation effects}

The analytic expression of the BFKL cross-section \eqref{mnjnll} lacks energy-momentum conservation, because these effects are formally higher-order corrections in this framework. However it has been argued \cite{dds,emc} that of the terms which conserve energy-momentum could be numerically important for phenomenological analysis. Therefore we shall estimate their magnitude for the observable \eqref{obs}. In order to do so, we will use the proposal of \cite{dds} which amounts to substitute $\Delta\eta$ in \eqref{mnjnll} by an effective rapidity interval
$y_{eff}.$ More advanced Monte Carlo approaches were later developed \cite{emc}, but we choose to stick to more insightful analytic calculations.

The effective rapidity is defined in the following way
\be
y_{eff}(p,Q,R,\Delta\eta,y)=\Delta\eta\
\lr{\int d\phi\cos(p\phi)\f{d\sigma^{O(\alpha_s^3)}}{d\Delta\eta dy dQ dR d\Delta\Phi}}^{-1}
\int d\phi\cos(p\phi)\f{d\sigma^{LL-BFKL}}{d\Delta\eta dy dQ dR d\Delta\Phi}
\ee
where $d\sigma^{O(\alpha_s^3)}$ is the exact $2\!\to\!3$ contribution to the
$hh\!\rightarrow\!JXJ$ cross-section at order $\alpha_s^3$ \cite{2to3}, and $d\sigma^{LL-BFKL}$ is the LL-BFKL result. One has $y_{eff}(\Delta\eta\!\to\infty\!)=\Delta\eta.$ In this way, when used in \eqref{newvar}, the expansion of the cross-section with respect to $\alpha_s$ is exact up to order $\alpha_s^3$ while the large $\Delta\eta$ limit is unchanged. To compute $d\sigma^{O(\alpha_s^3)},$ we used the standard jet cone size $R_{cut}\!=\!0.5$ when integrating over the third particle's momentum. The main feature of
$y_{eff}$ is that it is only slightly smaller than $\Delta\eta$ for $R\!=\!1,$ but that it decreases quickly with $R$ deviating from 1 \cite{dds}.

\begin{figure}[t]
\begin{center}
\epsfig{file=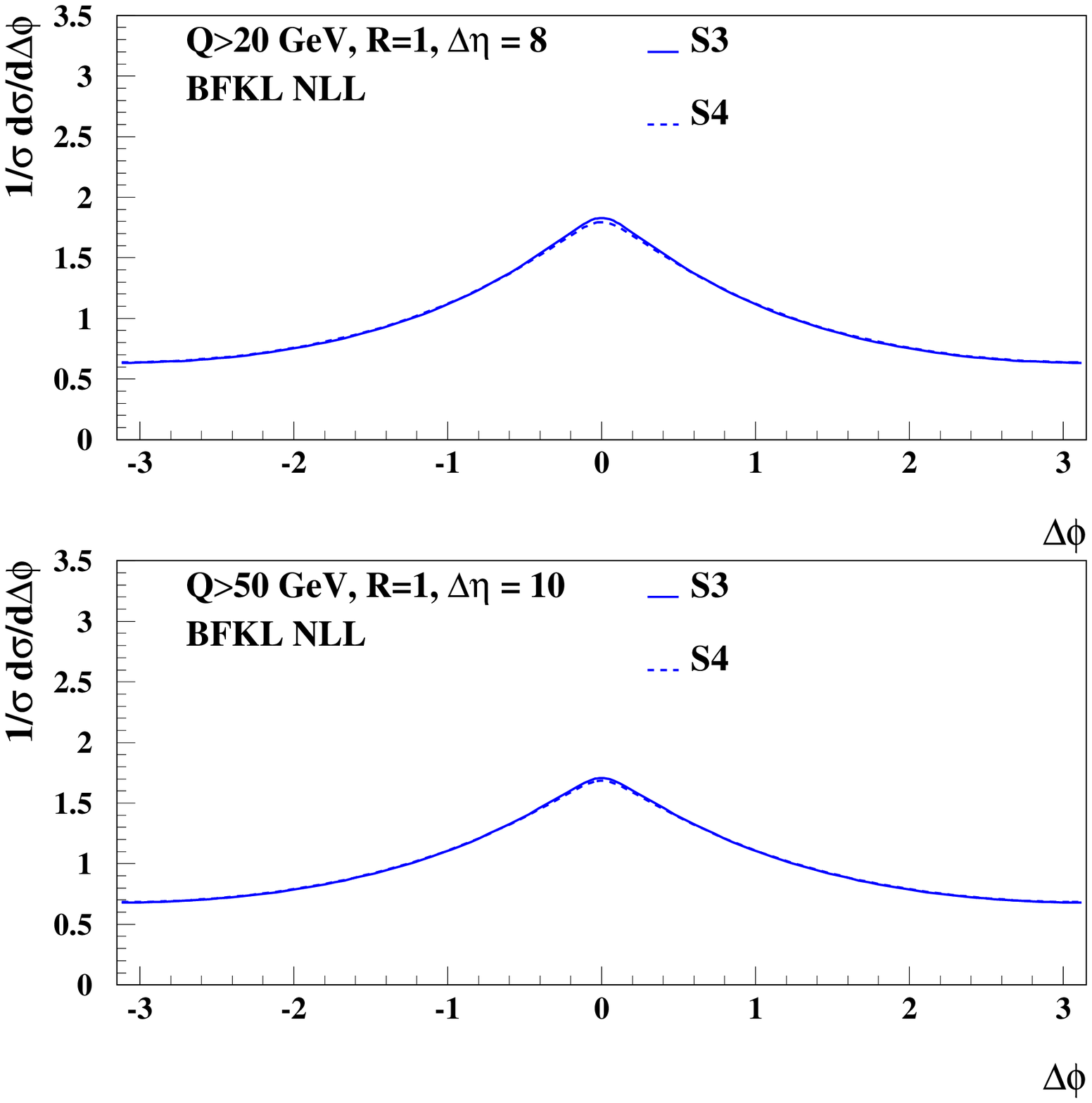,width=8.3cm}
\hfill
\epsfig{file=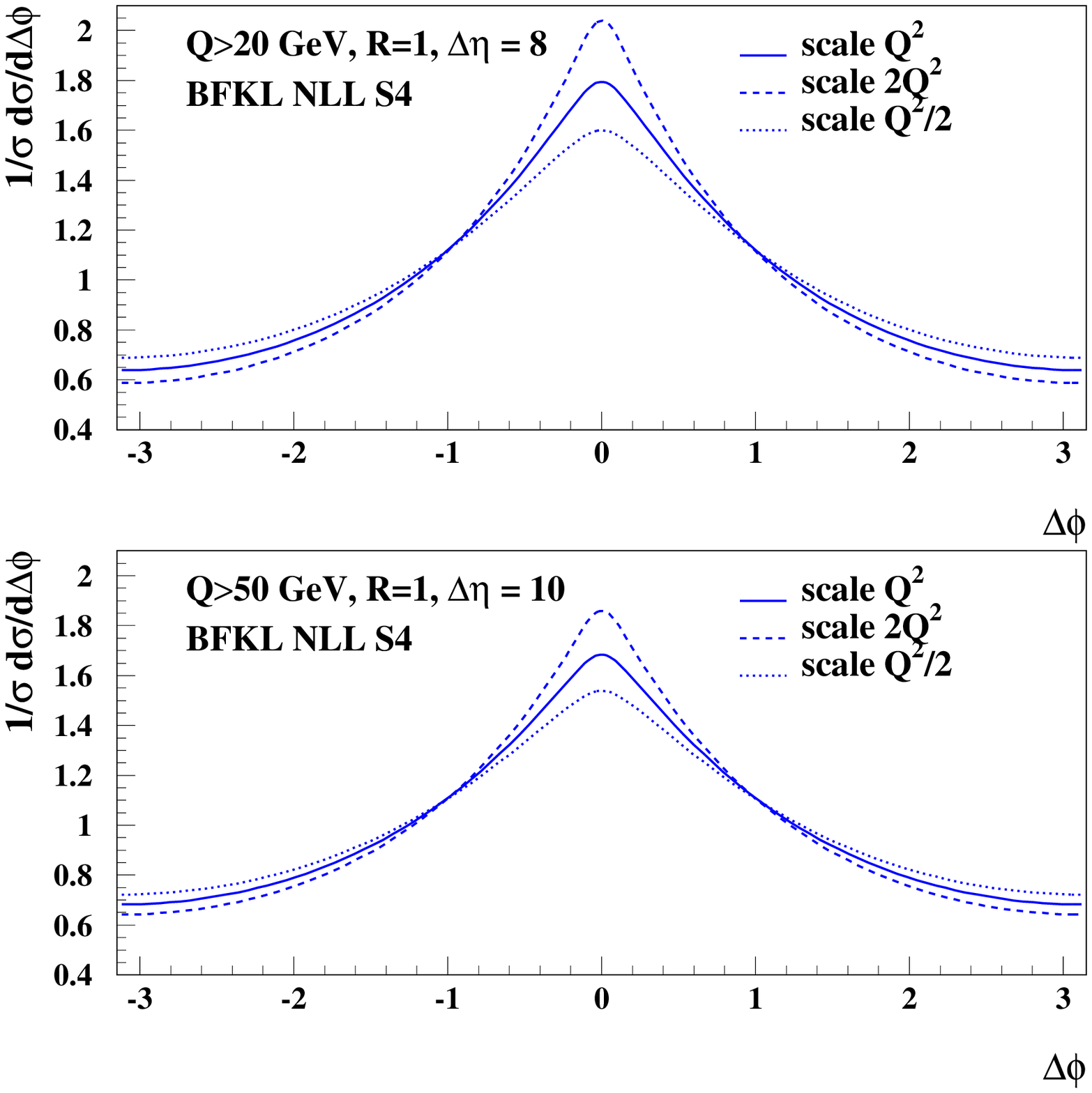,width=8.3cm}
\caption{Resumation-scheme and renormalization-scale dependencies of the Mueller-Navelet jet $\Delta\Phi$ distribution \eqref{obs} in the NLL-BFKL framework. Upper plots: $R\!=\!1,$ $\Delta\eta\!=\!8$ and Tevatron (run 2) kinematics; lower plots: 
$R\!=\!1,$ $\Delta\eta\!=\!10$ and LHC kinematics. The left plots show a comparison of the S3 and S4 schemes while the right plots display results obtained with the three renormalization scales $Q^2/2,\ Q^2,\ 2Q^2.$}
\end{center}
\end{figure}

\begin{figure}[t]
\begin{center}
\epsfig{file=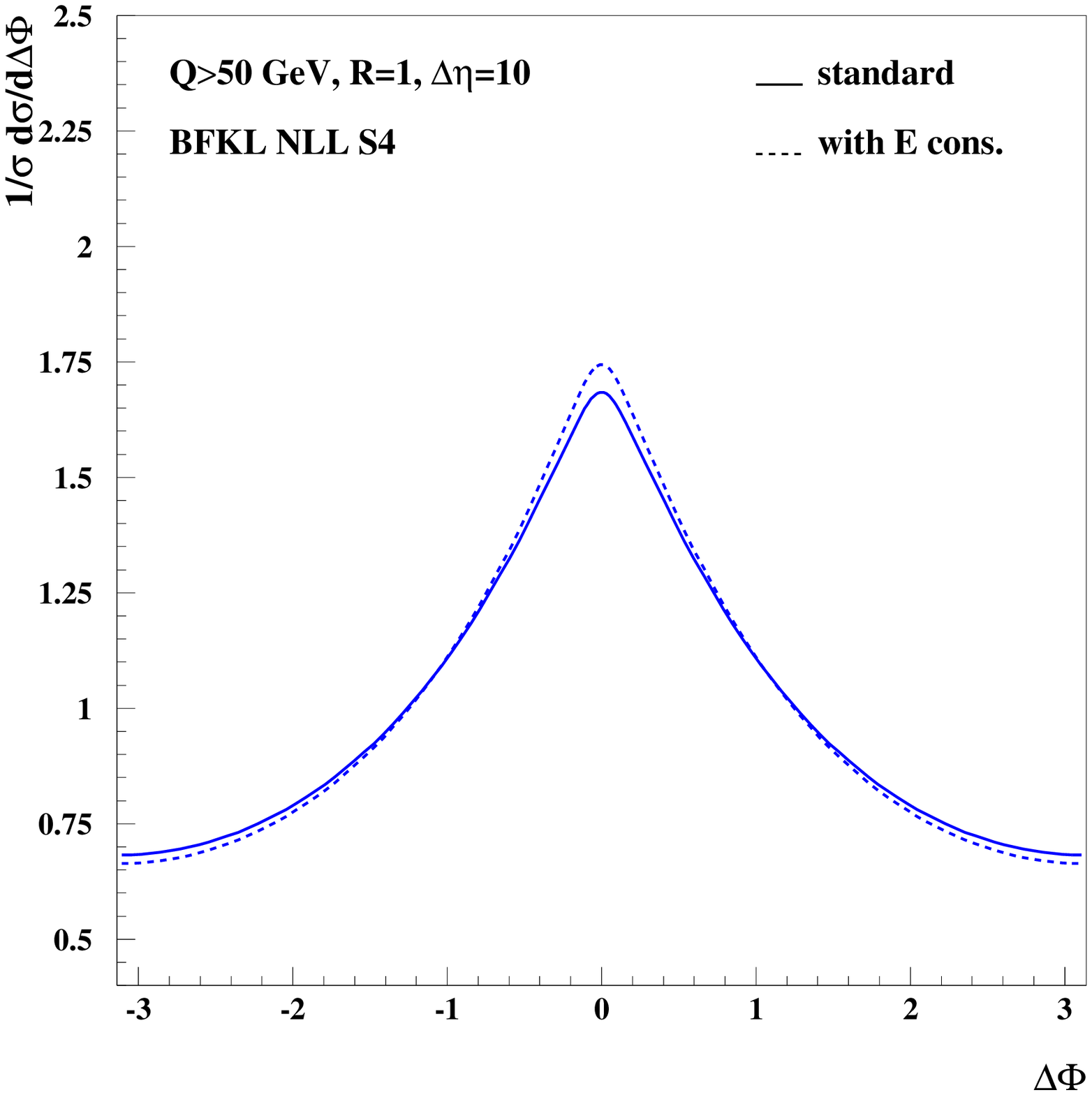,width=5.7cm}
\hfill
\epsfig{file=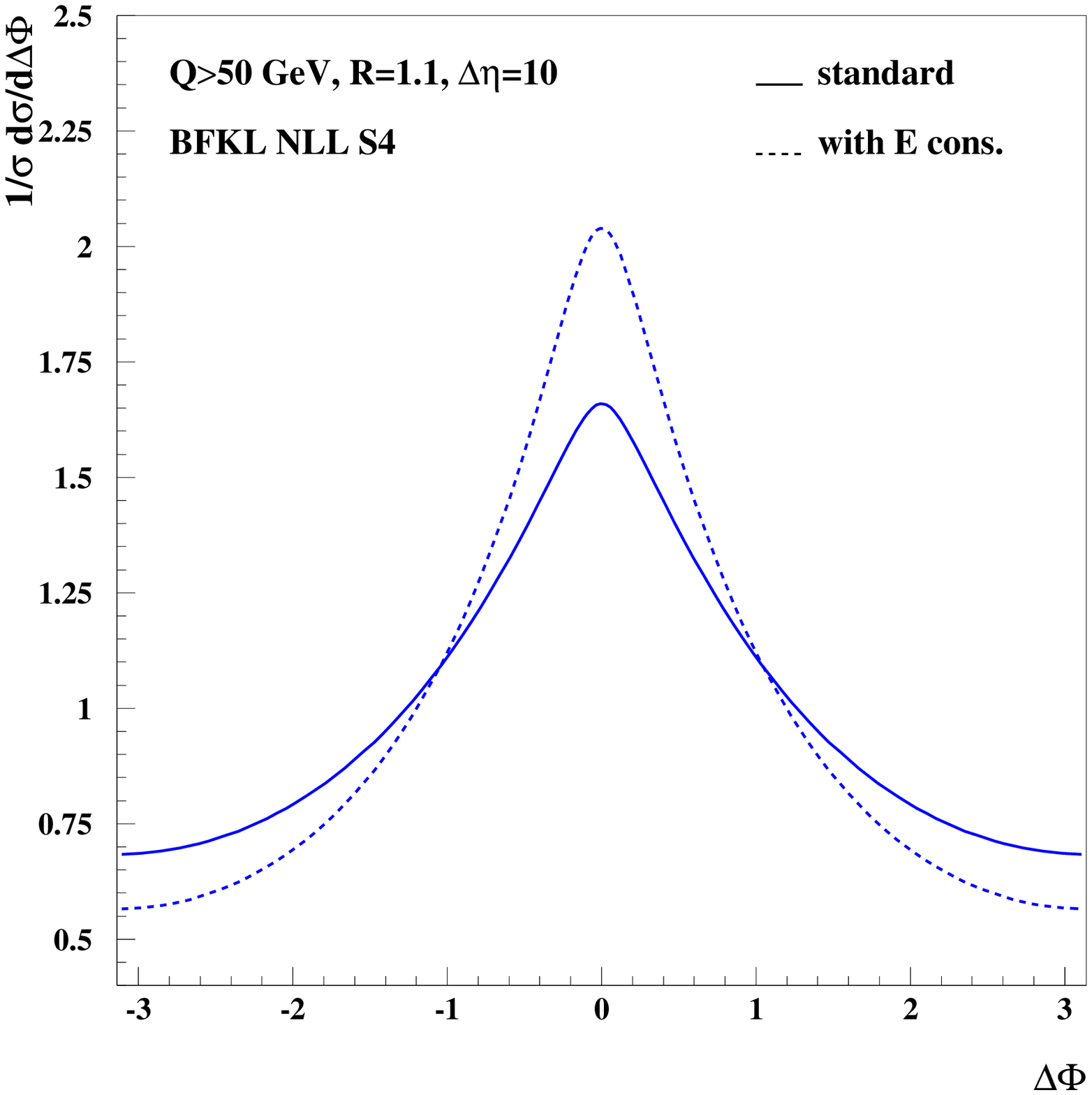,width=5.7cm}
\hfill
\epsfig{file=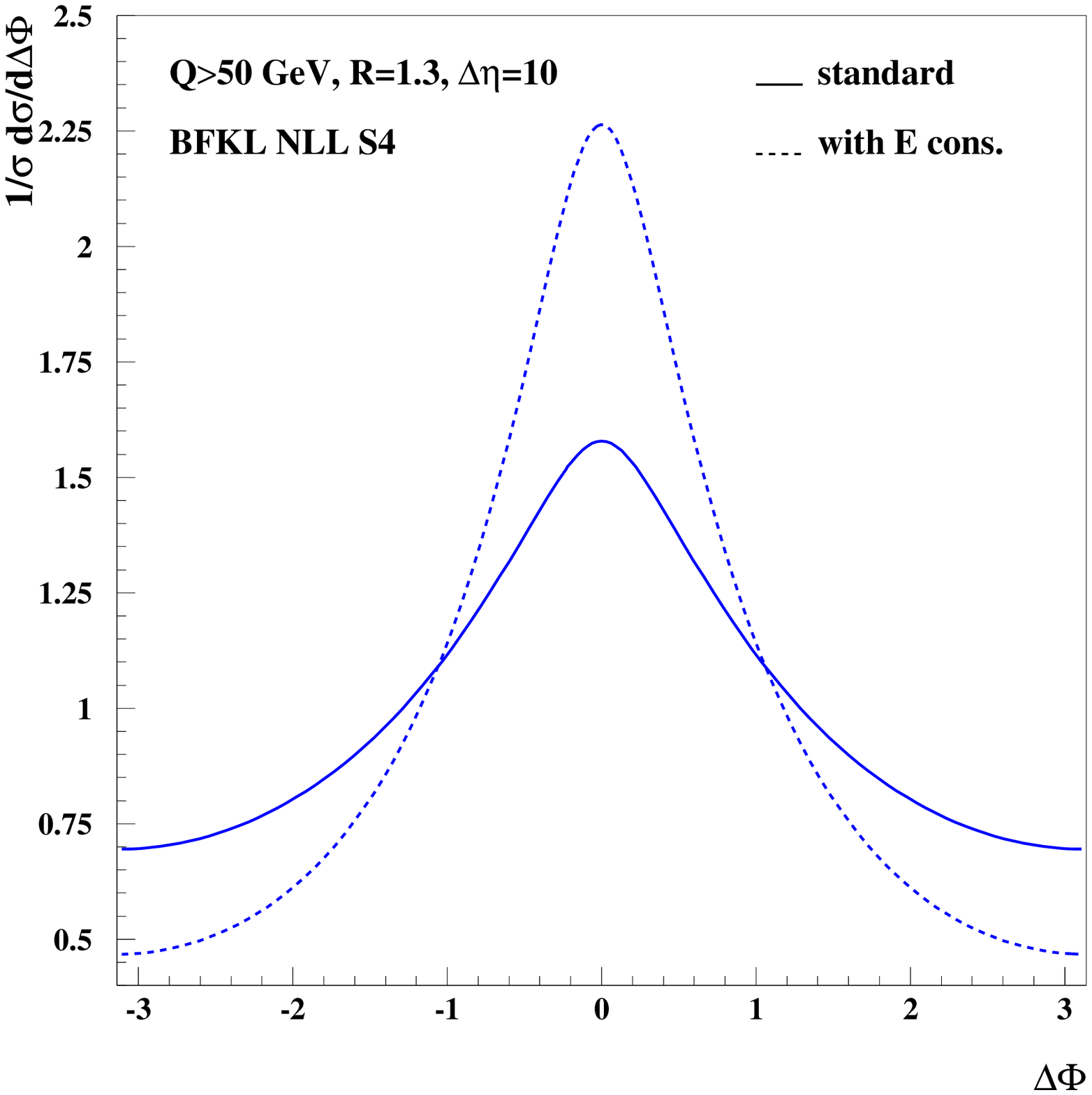,width=5.7cm}
\caption{Effects of energy conservation on the Mueller-Navelet jet $\Delta\Phi$ distribution for
$\Delta\eta=10$ and LHC kinematics. Left plot: $R=1;$ the effect is minimal. Central plot: $R=1.1,$ right plot: $R=1.3;$ the azimuthal correlation increases with $R$ deviating from 1 (instead of decresing) after energy-momentum conservation is included.}
\end{center}
\end{figure}

As shown in Fig.6, where the observable \eqref{obs} is plotted for LHC kinematics and
$\Delta\eta\!=\!10,$ this behavior is confirmed. Indeed, when $R\!=\!1$ the effect is minimal, the azimuthal correlation is only slightly bigger with energy momentum conservation. By contrast when $R\!\neq\!1$, the azimuthal correlation is much bigger with energy momentum conservation than without, and the effect is more and more important as $R$ deviates from 1. Therefore the modification of the $\Delta\phi$ spectrum with respect to $R$ is a measure of the role of energy-momentum conservation effects: without them the azimuthal correlation decreases with $R$ deviating from 1 while it is the opposite if such effects are included.

\subsection{Mueller-Navelet jets at CDF}

\begin{figure}[t]
\begin{center}
\epsfig{file=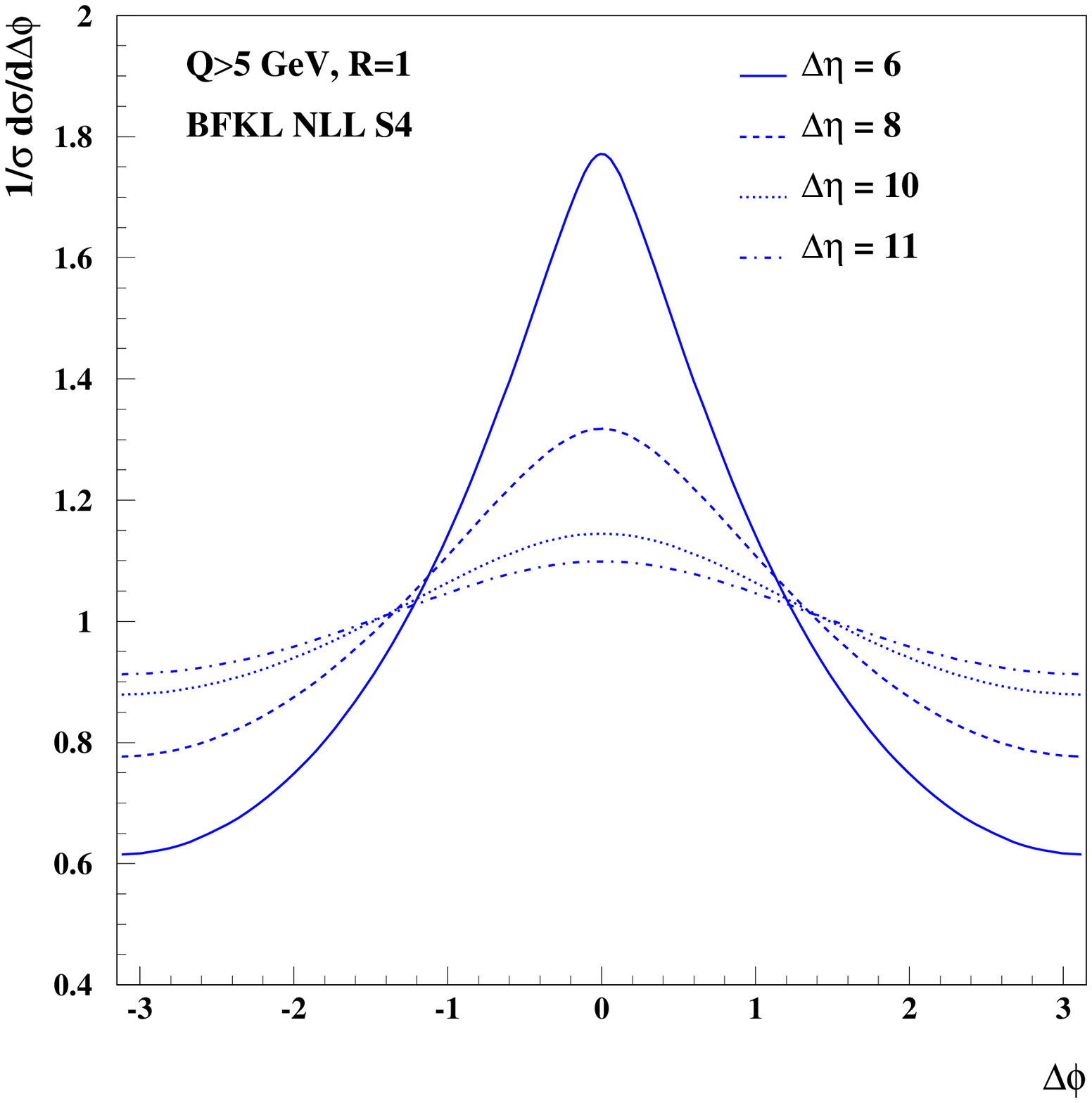,width=8cm}
\hfill
\epsfig{file=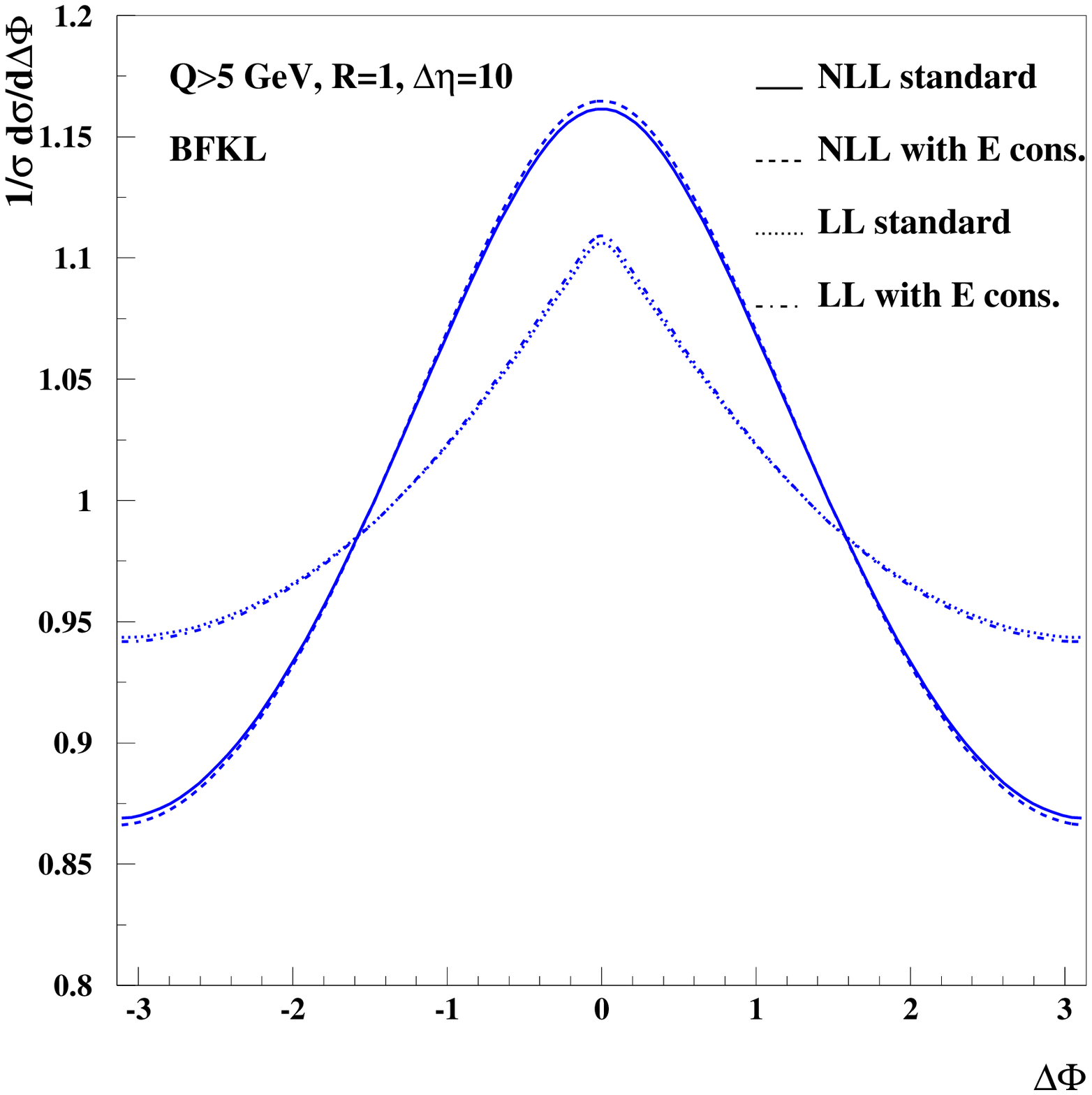,width=8cm}
\caption{The Mueller-Navelet jet $\Delta\Phi$ distribution \eqref{obs} for CDF kinematics and $R=1.$ Left plot: NLL-BFKL predictions for $\Delta\eta=6,\ 8,\ 10,\ 11.$ Right plot: comparison with the LL-BFKL result and calculations taking into account energy conservation, this effect is small as $R=1.$}
\end{center}
\end{figure}

The CDF collaboration recently installed detectors called Miniplugs in the forward and backward regions. These detectors allow to increase the acceptance in rapidity and transverse momentum to measure very forward jets. It will be possible to measure jets separated in rapidity by more than 10 units and with transverse momenta as low as $5\ \mbox{GeV}.$ It is also worth pointing out that while the CDF Miniplug detectors are not prefectly suited for energy measurements (the jet containment will be poor: the depth of the calorimeters is only one $\lambda$), they are especially interesting in the case of the observable studied here, which focuses on the difference in azimuthal angle between the jets.

The NLL-BFKL predictions for the Mueller-Navelet jet $\Delta\Phi$ distribution with CDF kinematics is represented in Fig.7. With such low values of transverse momenta and large values of rapidity interval between the two jets, it is also likely that saturation effects could play an important role. First estimations \cite{mnjsat} (obtained with less favorable kinematics) indicate so when considering saturation effets damping the LL-BFKL exponential growth. Studying saturation effects with NLL-BFKL growth certainly deserves more study. First steps have been taken in Ref.
\cite{nllsat}, but the problem of phenomenology for hadron colliders has yet to be addressed.

\section{Conclusion and Outlook}

We have investigated the decorrelation of Mueller-Navelet jets with respect to their relative azimuthal angle $\Delta\Phi$ in the BFKL framework at NLL accuracy. Using renormalization-group improved NLL kernels $\chi_{NLL}\lr{p,\g,\omega}$ in the S3 and S4 schemes, the NLL-BFKL effects were taken into account through an effective kernel obtained from the implicit equation
\eqref{eff}. This allowed our phenomenological study of NLL-BFKL effects in Mueller-Navelet jet production. Our present goal is to motivate future measurements at the Tevatron (Run 2) and at the LHC \cite{mnjcms}. A future comparison with the data will require to adapt our predictions to experimental cuts and perhaps to less differential cross sections.

The present study, devoted to the $\Delta\Phi$ spectrum \eqref{obs}, confirms the expectations 
that when increasing the rapidity interval between the jets $\Delta\eta,$ the decorrelation increases, and that NLL corrections decrease the azimuthal decorrelation with respect to the LL-BFKL results. We also investigated this effect as a function of $R\!=\!k_2/k_1,$ the ratio between the jets transverse momenta: when $R$ deviates from 1, the azimuthal decorrelation increases. Our predictions were obtained with standard expectations of Tevatron and LHC kinematical possibilities. However, we also presented predictions for the Mueller-Navelet jet $\Delta\Phi$ distribution having in mind the CDF forward detector which features a quite favorable kinematical reach ($Q\!>\!5\ GeV$ and $\Delta\eta\!>\!10$).

For the observable \eqref{obs}, we noticed that the differences between the different schemes are quite small, while the dependence on the choice of renormalization scale is of about 5 percent in general and reaches 20 percent around $\Delta\Phi\!=\!0.$ Energy-momentum conservation effects are minimal for $R\!=\!1,$ but they increase quite rapidly as $R$ deviates from 1. In fact, they reverse the trend discussed above: with energy-momentum conservation implemented, the azimuthal decorrelation decreases as $R$ deviates from 1.

With such low values of transverse momenta and large values of rapidity interval, Mueller-Navelet jet measurements would allow for a detailed study of the QCD dynamics of Mueller-Navelet jets, both for investigating fixed-order pQCD versus BFKL predictions, but also with respect to possible saturation effects. In these contexts, the measurement of the $\Delta\Phi$ integrated cross-section would be very interesting by itself, but a realistic phenomenological study should incorporate the next-to-leading order jet impact factors in the calculation. Indeed, their effect will not be suppressed as it likely is in the case of the normalized cross-section we have studied in this paper.

\begin{acknowledgments}

We would like to thank Robi Peschanski for commenting on the manuscript. C.M. is supported in part by RIKEN, Brookhaven National Laboratory and the U.S. Department of Energy [DE-AC02-98CH10886].

\end{acknowledgments}
\section*{Appendix: The S3 and S4 schemes for non-zero conformal spins}

In this Appendix, we show how to extend the regularisation procedure of \cite{salam} to non zero conformal spins $p\neq 0$. We obtain $\chi_{NLL}\lr{p,\g,\omega}$ for the S3 and S4 schemes (recently two preprints appeared where the S3 scheme \cite{sabsch} and the other Salam schemes \cite{schwen} have also been extended).

The starting point is the scale invariant (and $\g\!\leftrightarrow\!1-\g$ symmetic) part of the NLL-BFKL kernel
\bea
\chi_1(p,\g)=\f32\zeta(3)+\lr{\f{1+5b}3-\f{\zeta(2)}2}\chi_{LL}(p,\g)-\f{b}2\chi_{LL}^2(p,\g)
+\f14\left[\psi''\lr{\g+\f{p}2}+\psi''\lr{1-\g+\f{p}2}\right]\nn\\
-\f12\left[\phi(p,\g)+\phi(p,1-\g)\right]
-\f{\pi^2\cos(\pi\g)}{4\sin^2(\pi\g)(1-2\g)}
\left\{\left[3+\lr{1+\f{N_f}{N_c^3}}\f{2+3\g(1-\g)}{(3-2\g)(1+2\g)}\right]\delta_{0p}
\right.\nn\\\left.-\lr{1+\f{N_f}{N_c^3}}\f{\g(1-\g)}{2(3-2\g)(1+2\g)}\delta_{2p}\right\}\eea
with $b$ given in \eqref{runc}, $\chi_{LL}$ given in \eqref{chill}, and 
\bea
\phi(p,\g)=\sum_{k=0}^\infty\f{(-1)^k}{k+\g+p/2}\left\{
\psi'(k+1)-\psi'(k+p+1)+\f{\psi(k+p+1)-\psi(k+1)}{k+\g+p/2}\right.\nn\\\left.
+\f{(-1)^k}4\left[\psi'\lr{\f{k+p+2}2}
-\psi'\lr{\f{k+p+1}2}+\psi'\lr{\f{k+2}2}-\psi'\lr{\f{k+1}2}\right]\right\}\ .
\label{klcor}\eea
Note that for the terms on the first line of \eqref{klcor} inside the curly brakets, we have corrected the signs with respect to Ref. \cite{kotlip}, where they are misprinted (the signs are correct in Ref. \cite{lipkot}). As is the case for 
$\chi_{LL}(p,\g),$ the kernel $\chi_1(p,\g)$ has poles at $\g=-p/2$ and $\g=1+p/2.$ The pole structure at $\g=-p/2$ (and by symmetry at $\g=1+p/2$) is:
\be
\chi_1(p,\g)=-\f1{2\lr{\g+\f{p}2}^3}+\f{d_2(p)}{\lr{\g+\f{p}2}^2}
+\f{d_1(p)}{\lr{\g+\f{p}2}}+{\cal O}(1)
\ee
with
\bea
d_1(p)=\f{1+5b}3-\f{\pi^2}8+b[\psi(p+1)-\psi(1)]+
\f18\left[\psi'\lr{\f{p+1}2}-\psi'\lr{\f{p+2}2}+4\psi'\lr{p+1}\right]
\nn\\-\lr{67+13\f{N_f}{N_c^3}}\f{\delta_{0p}}{36}
-\lr{1+\f{N_f}{N_c^3}}\f{47\delta_{2p}}{1800}
\eea
and
\be
d_2(p)=-\f{b}2-\f12[\psi(p+1)-\psi(1)]
-\lr{11+2\f{N_f}{N_c^3}}\f{\delta_{0p}}{12}-\lr{1+\f{N_f}{N_c^3}}\f{\delta_{2p}}{60}\ .
\ee
Note that $\chi_1(2,\g)$ also has a pole at $\g=0$ with residue $(1+N_f/N_c^3)/24.$ This manifestation of the non-analyticity \cite{kotlip} of $\chi_1(p,\g)$ with respect to the conformal spin does not alter the stability of the NLL prediction and a careful treatment of this singularity is not required.

\subsection{Extension of the S3 scheme}

The S3-scheme kernel $\chi_{S3}\lr{p,\g,\omega}$ is given by 
\bea
\chi_{S3}(p,\g,\omega)=[1-\bar\alpha A(p)]
\left[2\psi(1)
-\psi\lr{\g+\f{p+2\bar\alpha B(p)+\omega}2}
-\psi\lr{1-\g+\f{p+2\bar\alpha B(p)+\omega}2}\right]
\nn\\+\bar\alpha\left\{\chi_1(p,\g)+A(p)\chi_{LL}(p,\g)
+\lr{B(p)+\f{\chi_{LL}(p,\g)}2}\left[\psi'\lr{\g+\f{p}2}+\psi'\lr{1-\g+\f{p}2}\right]\right\}
\eea
with $A(p)$ and $B(p)$ chosen to cancel the singularities of $\chi_1(p,\g)$ at 
$\g=-p/2:$
\be
A(p)=-d_1(p)-\psi'(p+1)\ ,\hspace{1cm}B(p)=-d_2(p)+\f12[\psi(p+1)-\psi(1)]\ .
\ee

\subsection{Extention of the S4 scheme}

The S4-scheme kernel $\chi_{S4}\lr{p,\g,\omega}$ is given by 
\bea
\chi_{S4}(p,\g,\omega)=\chi_{LL}(p,\g)-f(p,\g)
+[1-\bar\alpha A(p)]f(p+\omega+2\bar\alpha B(p),\g)
\nn\\+\bar\alpha\left\{\chi_1(p,\g)+A(p)f(p,\g)
+\lr{B(p)+\f{\chi_{LL}(p,\g)}2}
\left[\lr{\g+\f{p}2}^{-2}+\lr{1-\g+\f{p}2}^{-2}\right]\right\}
\eea
with
\be
f(p,\g)=\f1{\g+\f{p}2}+\f1{1-\g+\f{p}2}\ .
\ee
In this scheme, $A(p)$ and $B(p)$ are given by:
\be
A(p)=-d_1(p)-\f12\left[\psi'(p+1)-\psi'(1)+\f1{(p+1)^2}\right]
\ ,\hspace{1cm}B(p)=-d_2(p)+\f12[\psi(p+1)-\psi(1)]\ .
\ee


\end{document}